\documentclass{DISproc}

\newcommand{\dphi}{\Delta\phi\xspace}
\newcommand{\pizero}{\mbox{$\pi^{0}$}\xspace}

\newcommand{\dAu}{$d+$Au\xspace}
\newcommand{\pp}{$p$+$p$\xspace}
\newcommand{\rg}{\mbox{$R_g^{Au}$}\xspace}

\newcommand{\rda}{\mbox{$R_{\rm dA}$}\xspace}
\newcommand{\jda}{\mbox{$J_{\rm dA}$}\xspace}

\newcommand{\meanNcoll}{\mbox{$\langle N_{\rm coll} \rangle$}\xspace}

\newcommand{\sqsn}{\mbox{$\sqrt{s_{_{NN}}}$}\xspace}
\newcommand{\pT}{\mbox{$p_T$}\xspace}
\newcommand{\xfrag}{$x_{Au}^{frag}$\xspace}

\begin{document}
\title{Probing the Low-x Structure of the Nucleus with the PHENIX Detector}

\author{{\slshape Mickey Chiu}\\[1ex]
Brookhaven National Lab, Upton, NY 11973 USA\\
}

\contribID{133}

\doi  

\maketitle

\begin{abstract}
One of the fundamental goals of the PHENIX experiment is to understand
the structure of cold nuclear matter, since this serves as the initial
state for heavy-ion collisions. Knowing the initial state is vital for
interpreting measurements from heavy-ion collisions. Moreover, the
structure of the cold nucleus by itself is interesting since it is a
test-bed for our understanding of QCD.  In particular there is the
possibility of novel QCD effects such as gluon saturation at low-x
in the nucleus. At RHIC we can probe the behavior of gluons at
low-x by measuring the pair cross-section of di-hadrons from di-jets
in d+Au collisions.  Our results show a systematic decrease in the pair
cross-section as one goes to smaller impact parameters of the nucleus,
and also as one goes to lower Bjorken x.  There is a possibility that these
interesting effects come from gluon recombination at low x in the Au nucleus. 
\end{abstract}

\section{Introduction}

Deuteron-gold collisions at RHIC provide a means to
explore nuclear effects on the initial-state parton
densities in the nucleus, which is vitally important
to understanding the baseline production for Quark-Gluon Plasma
studies in heavy-ion
collisions.  RHIC experiments have shown that 
single inclusive hadron yields in the forward (deuteron)
rapidity direction for $\sqsn = 200$ GeV \dAu
collisions are suppressed relative to \pp
collisions~\cite{rda_brahms, rda_star, rcp_phenix}. The mechanism
for the suppression has not been firmly established. 
Many effects have been proposed for this suppression, such as 
gluon saturation~\cite{cgc, monojets}, initial state energy
loss~\cite{vitev2, strikman}, parton recombination~\cite{hwa},
multi-parton interactions~\cite{strikman_parton}, and
leading and higher-twist shadowing~\cite{GSV, vitev}.

One set of measurements that might help to distinguish between the
competing models is forward azimuthally correlated di-hadron correlation
functions, which directly probe di-jet production through their
2$\rightarrow$2 back-to-back peak at $\dphi = \pi$.  This
technique has been used extensively at RHIC and is described in detail
elsewhere~\cite{ida_central_phenix, ida_phenix,jda_paper}.  
The di-hadron results presented here were obtained from \pp and \dAu runs
in 2008 with the PHENIX detector and include a new electromagnetic calorimeter, the Muon Piston
Calorimeter (MPC), with an acceptance of $3.1 < \eta < 3.8$ in pseudorapidity
and $0 < \phi < 2 \pi$.  

Di-hadron measurements can probe more precise ranges of parton $x$ in a gold nucleus
than do single hadron probes (e.g., \rda). At forward rapidities, a single
hadron probe will cover a very broad range of x, $10^{-3}<x_{Au}<0.5$, thus mixing together
the shadowing, anti-shadowing, and even EMC effects~\cite{GSV}.  Azimuthally
correlated di-hadron measurements also enhance the di-jet 
fraction in the event selection, since one selects only the back-to-back hadrons.

By performing several correlation measurements with
particles at different \pT and rapidities, one can systematically scan different
$x$ ranges with an observable that is enhanced for the leading-order perturbative
QCD component.  Probing the $x$ dependence of the effect is an important test
since most models predict that any effects
should be stronger at smaller $x$. Particles at higher pseudorapidities are produced
from smaller $x$, so measuring hadrons from more forward rapidities
should probe smaller x.

\section{PHENIX MPC \dAu di-Hadron Correlations}

For this analysis, back-to-back \pizero-\pizero or hadron-\pizero pairs are measured
with one particle at mid-rapidity, and the other at forward rapidity.  Back-to-back
cluster-\pizero pairs are also measured where both are in the forward rapidity region.
The clusters are reconstructed from the energy deposit of photons in the MPC,
and are estimated to be at least 80\% dominated by \pizero's, with the remainder coming
from single photons from decays of $\eta$'s and from direct photons.  Further details of the
analysis are available in~\cite{jda_paper}.

\begin{wrapfigure}{l}{0.65\textwidth}
  \includegraphics[width=0.6\textwidth]{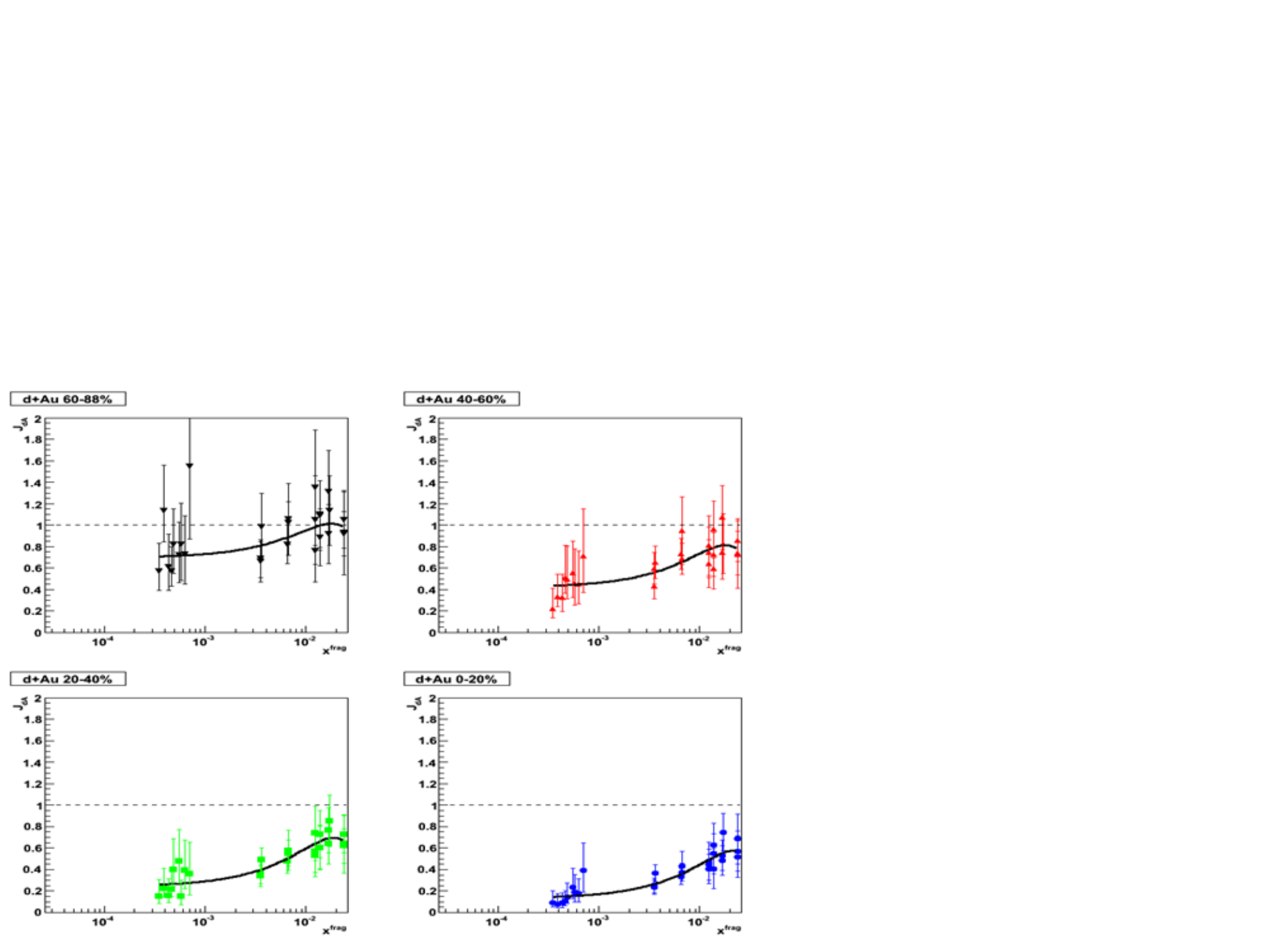}
  \caption{\jda vs \xfrag for 4 different centrality bins. The points are taken from ~\cite{jda_paper}
  and fit with the same parametric function as those used for the EPS09 set of nuclear pdf's~\cite{eps09}.
  The errors are the statistical and systematic errors from ~\cite{jda_paper} added in quadrature.}
  \label{Fig:JDA_EPS09}
\end{wrapfigure}

From the pairs we extract the relative yield, \jda, of correlated
back-to-back hadrons in \dAu collisions compared to \pp collisions scaled with the average
number of binary nucleon collisions $\langle N_{coll} \rangle$, where 
\begin{equation*}
\jda = \frac{1}{\langle N_{coll} \rangle}\frac{\sigma_{dA}^{pair}/\sigma_{dA}}{\sigma_{pp}^{pair}/\sigma_{pp}}
\end{equation*}
and is explained in detail in \cite{jda_orig}.  
\jda is simply the analog of the usual nuclear modification factor \rda but for hadron pairs.
The $\sigma_{dA,pp}$ are the \pp or \dAu inelastic cross-sections, while $\sigma^{pair}_{dA,pp}$ are the cross-sections for 
di-hadron pair production, and is used as a proxy for di-jets in PHENIX.

In Fig. \ref{Fig:JDA_EPS09}, we have plotted the values of \jda 
versus \xfrag for four different \dAu centrality selections.  
\xfrag is defined as
\begin{equation*}
x_{Au}^{frag}= (\langle p_{T1}\rangle e^{-\langle \eta_1\rangle} + 
\langle p_{T2}\rangle e^{-\langle \eta_2\rangle})/\sqrt{s_{NN}}
\end{equation*}
and can be directly measured experimentally.  \xfrag should be
correlated with the Bjorken x that is probed in the nucleus,
assuming that a normal leading order
(LO) perturbative QCD framework applies for this data.
In the case of 2$\rightarrow$2 LO processes,  
the variable  $x_{Au}^{frag}$ is lower than $x_{Au}$ by the mean 
fragmentation fraction, $\langle z \rangle$, of the struck parton 
in the Au nucleus. 
From the plot, one can see that \jda decreases with increasing centrality,
or equivalently with increasing nuclear
thickness.  The suppression also increases as one goes to lower
\xfrag in the nucleus probed by the deuteron. 

In Fig. \ref{Fig:JDA_CentDep}, the \jda values for three different \xfrag
are plotted versus \meanNcoll, the mean number of binary collisions, in
the four centrality classes depicted in Fig. \ref{Fig:JDA_EPS09}.  One
can clearly see from this plot a systematic decrease of \jda with greater \meanNcoll,
as well as with decreasing x.  The decrease is approximately linear.

\section{Discussion}

In a leading order pQCD picture, the variable \jda is
\begin{equation}
J_{dA} = \frac{\sigma_{dA}^{pair}/\sigma_{dA}}{\langle N_{coll} \rangle\, \sigma_{pp}^{pair}/\sigma_{pp}} \approx
\frac{f^a_d(x^a_d)\otimes f^b_{Au}(x^b_{Au})\otimes\hat{\sigma}^{ab\rightarrow cd}\otimes\mathcal{D}(z_c,z_d)}
{\langle N_{coll} \rangle \, f^a_p(x^a_p)\otimes f^b_p(x^b_p)\otimes\hat{\sigma}^{ab \rightarrow cd}\otimes\mathcal{D}(z_c,z_d)}
\end{equation}
for partons a+b going to outgoing jets c+d, which then fragment to hadrons with
longitudinal fractions $z_c$, $z_d$.  
In the above convolutions over the parton distribution functions ($f$), the parton-parton
cross-section $\hat{\sigma}$, and fragmentation functions $\mathcal{D}$, most of the terms
are expected to be roughly similar between p+p and d+Au except for the nuclear gluon 
parton distribution (pdf).  

\begin{wrapfigure}{l}{0.5\textwidth}
  \begin{center}
  \includegraphics[width=0.45\textwidth]{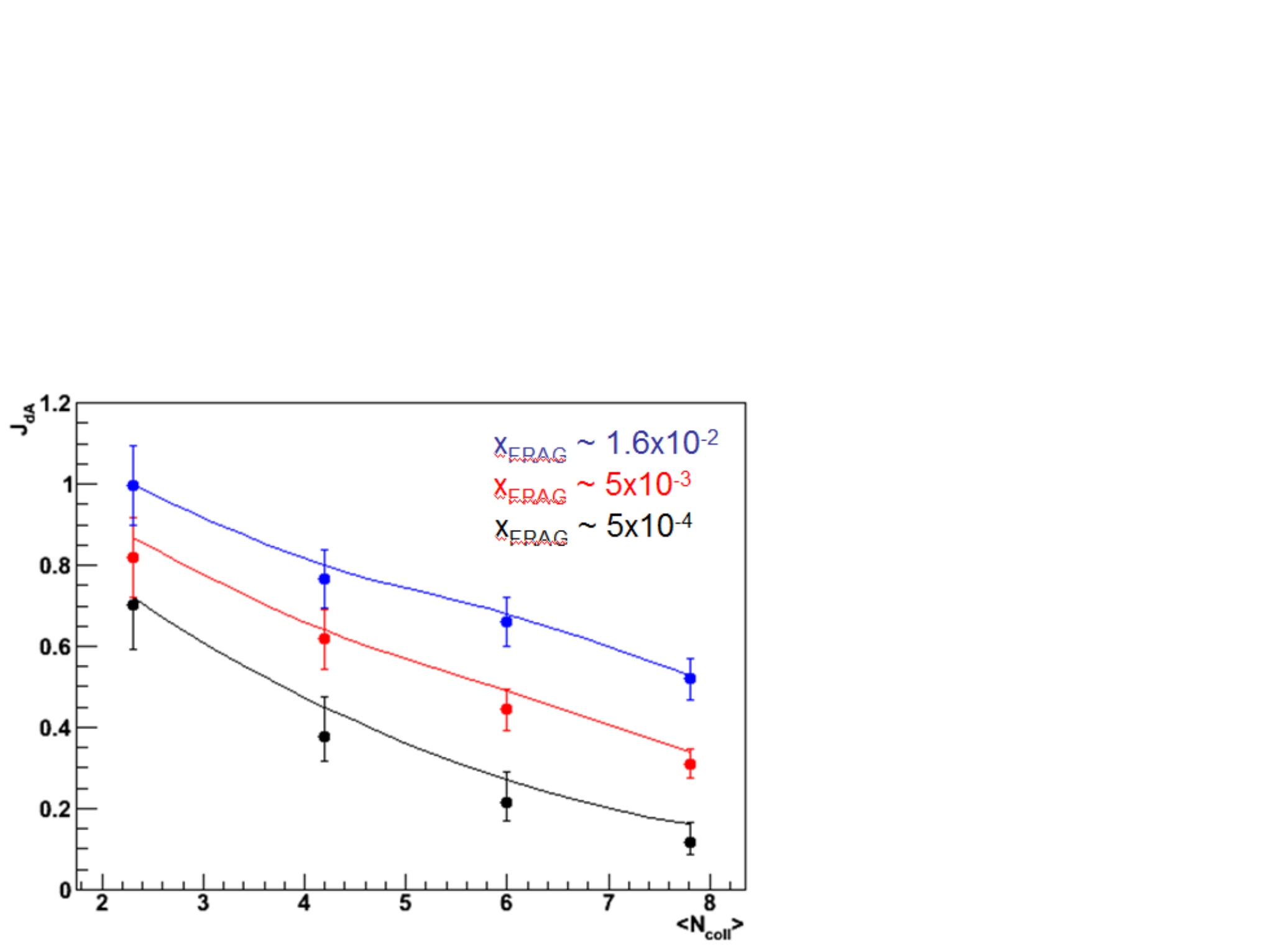}
  \caption{\jda vs \meanNcoll, the mean number of binary collisions,
  for different \xfrag values.}
  \label{Fig:JDA_CentDep}
  \end{center}
\end{wrapfigure}

\begin{figure}[htbp]
  \begin{center}
  \includegraphics[width=0.45\textwidth]{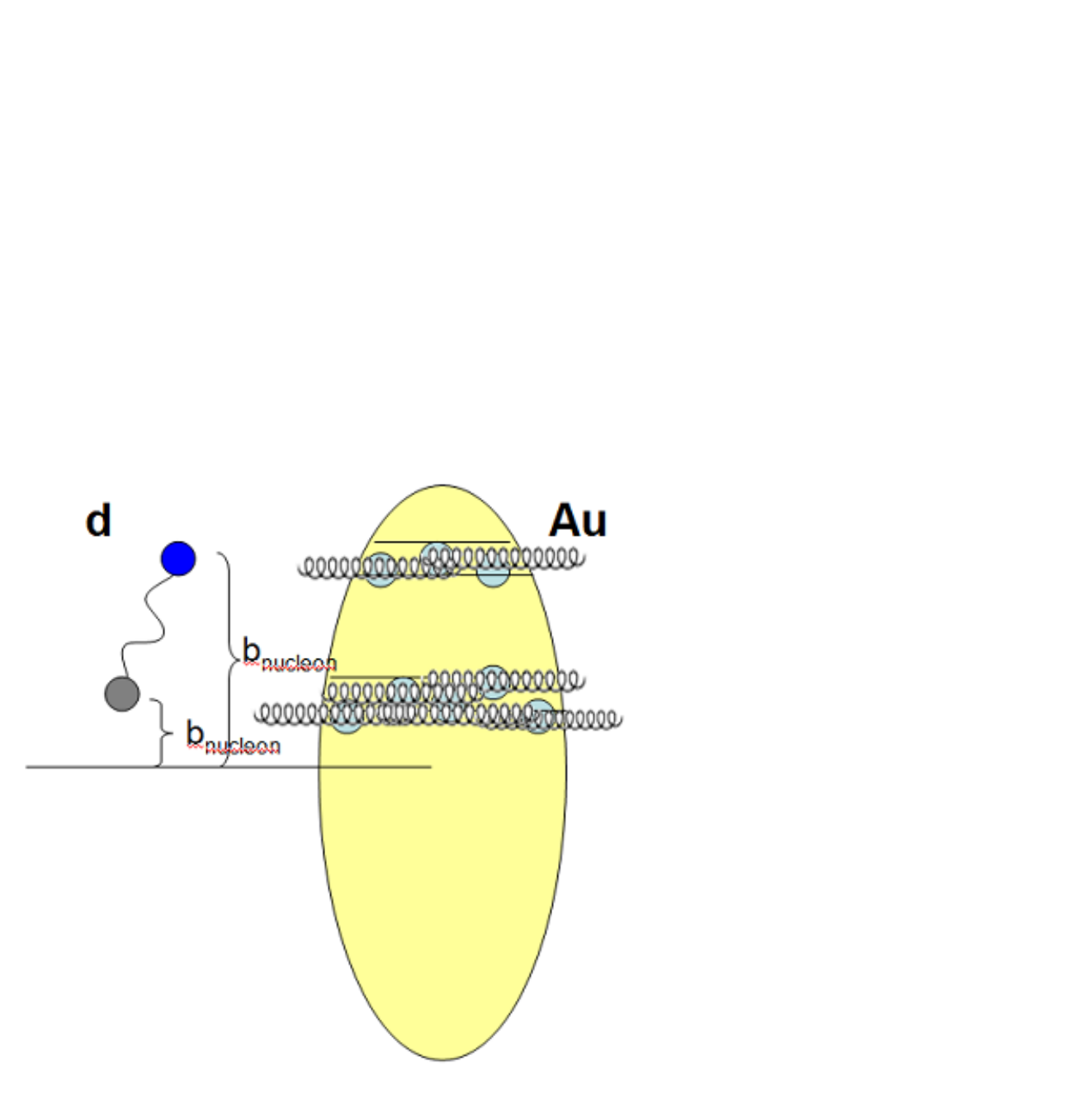}
  \caption{Schematic illustration depicting the increasing overlapping of gluons with smaller impact parameter
   due to the increasing number of nucleons along a straight line through the nucleus. This increases the
   probability for gluon recombination effects.}
  \label{Fig:GluonOverlap}
  \end{center}
\end{figure}

Naively, \jda might be largely dominated by the modification to the nuclear gluon
pdf, since most of the events with di-hadrons at forward rapidities consist of
a high-x parton from the deuteron and a low-x gluon from the gold nucleus.
Assuming this to be true, one can then associate \jda with the relative modification of
the nuclear gluon distribution, \rg, i.e.,
\begin{equation*}
\jda \sim \rg = G_{Au}(x,Q^2)/A\,G_p(x,Q^2)
\end{equation*}
One can then interpret Fig. \ref{Fig:JDA_CentDep} as a systematic decrease
in the gluon distribution when one goes to the thicker parts of the nucleus,
perhaps due to recombination of the gluons, and that the recombination
creates a proportional decrease in the number of gluons with increasing
number of nucleons along a line in the nucleus.  This is schematically illustrated
in Fig. \ref{Fig:GluonOverlap}.  This decrease is stronger at lower x, which one
might expect since the transverse size of the gluons are larger for lower x.

If nature is kind and this data can be interpreted in terms
of a simple pQCD picture, then this data may provide 
valuable information on
how gluons recombine in the nucleus as a function of the thickness of the nucleus
and Bjorken x of the gluon.  Furthermore, it may be possible
to extract \rg, which is extremely important for understanding the quark gluon
plasma since it forms the main ingredient for production in heavy ion collisions.
However, even in a simple pQCD picture there have been proposed new effects in 
\dAu collisions, such as initial state energy loss~\cite{vitev2,strikman}, multi-parton
interactions~\cite{strikman_parton}, which may complicate the interpretation of this data.
Further theoretical studies are needed before a robust extraction of \rg is credibly
achieved.

\section{Bibliography}


{\raggedright
\begin{footnotesize}




\begin{thebibliography}{99}
\bibitem{rda_brahms}
  I.~Arsene {\it et al.}[BRAHMS Collaboration],
  Phys.\ Rev.\ Lett.\  {\bf 93} (2004) 242303
\bibitem{rda_star}
  J.~Adams {\it et al.}[Star Collaboration],
  Phys.\ Rev.\ Lett.\  {\bf 97} (2006) 152302
\bibitem{rcp_phenix}
  S.~Adler {\it et al.}[PHENIX Collaboration],
  Phys.\ Rev.\ Lett.\  {\bf 94} (2005) 082302
\bibitem{cgc}
  L.~D.~McLerran and R.~Venugopalan,
  Phys.\ Rev.\  D {\bf 49} (1994) 3352
\bibitem{monojets}
  D.~Kharzeev, E.~Levin and L.~McLerran,
  Nucl.\ Phys.\  A {\bf 748} (2005) 627
\bibitem{vitev2}
  I.~Vitev
  Phys.\ Rev.\  C {\bf 75} (2007) 064906
\bibitem{strikman}
  L.~Frankfurt and M.~Strikman,
  Phys.\ Lett.\  B {\bf 645} (2007) 412
\bibitem{hwa}
  R.~C.~Hwa, C.~B.~Yang and R.~J.~Fries,
  Phys.\ Rev.\  C {\bf 71} (2005) 024902
\bibitem{strikman_parton}
  M.~Strikman and W.~Vogelsang,
  Phys.\ Rev.\  D {\bf 83}, 034029 (2011)
\bibitem{GSV}
  V.~Guzey, M.~Strikman and W.~Vogelsang,
  Phys.\ Lett.\  B {\bf 603}, 173 (2004)
\bibitem{vitev}
  J.~w.~Qiu and I.~Vitev,
  Phys.\ Lett.\  B {\bf 632} (2006) 507
\bibitem{ida_central_phenix}
  S.~Adler {\it et al.}[PHENIX Collaboration],
  Phys.\ Rev.\ C.\  {\bf 73} (2006) 054903
\bibitem{ida_phenix}
  S.~Adler {\it et al.}[PHENIX Collaboration],
  Phys.\ Rev.\ Lett.\  {\bf 96} (2006) 222301
\bibitem{jda_paper}
  A.~Adare {\it et al.}  [PHENIX Collaboration],
  Phys.\ Rev.\ Lett.\  {\bf 107}, 172301 (2011)
\bibitem{jda_orig}
  A.~Adare {\it et al.}[PHENIX Collaboration],
  Phys.\ Rev.\ C.\  {\bf 78} (2008) 014901
\bibitem{eps09}
  K.~J.~Eskola, H.~Paukkunen and C.~A.~Salgado,
  JHEP {\bf 0904} (2009) 065
\end{thebibliography}
\end{footnotesize}
}


\end{document}